\documentstyle[aps,preprint,tighten,floats,epsf,rotate]{revtex}
\begin{document}
\draft

%
\title{Baryogenesis via Density Fluctuations with a Second Order 
Electroweak Phase Transition}
\author{Biswanath Layek \footnote{e-mail: layek@iopb.res.in},
Soma Sanyal \footnote{e-mail: sanyal@iopb.res.in},
and Ajit M. Srivastava \footnote{e-mail: ajit@iopb.res.in}}
\address{Institute of Physics, Sachivalaya Marg, Bhubaneswar 751005, 
India}
%
%
\maketitle
\widetext
\parshape=1 0.75in 5.5in
\begin{abstract}
We consider the presence of cosmic string induced density fluctuations 
in the universe at temperatures below the electroweak phase transition 
temperature. Resulting temperature fluctuations can restore the 
electroweak symmetry locally, depending on the amplitude of
fluctuations and the background temperature. The symmetry will be 
spontaneously broken again in a given fluctuation region as the
temperature drops there (for fluctuations with length scales smaller 
than the horizon), resulting in the production of baryon asymmetry. 
The time scale of the transition will be governed by the 
wavelength of fluctuation and, hence, can be much smaller than 
the Hubble time. This leads to strong enhancement in the production 
of baryon asymmetry for a second order electroweak phase
transition as compared to the case when transition happens due to
the cooling of the universe via expansion. For a two-Higgs  extension
of the Standard Model (with appropriate CP violation), 
we show that one can get the required 
baryon to entropy ratio if fluctuations propagate without getting
significantly damped. If fluctuations are damped rapidly, then
a volume factor suppresses the baryon production.
Still, the short scale of the fluctuation leads to enhancement
of the baryon to entropy ratio by at least 3 - 4 orders of magnitude 
compared to the conventional case of second order transition where
the cooling happens due to expansion of the universe.

\end{abstract}
\vskip 0.125 in
\parshape=1 0.75in 5.5in
\narrowtext

\section{Introduction}

 Baryogenesis at the electroweak phase transition in the universe is one 
of the most attractive models of solving the matter-antimatter problem of 
the universe \cite{ewb}. As compared to other models such as those based on
Grand Unified Theories (GUT), primordial black holes, etc., much 
of the physics of electroweak theory is well understood and is accessible to
laboratory experiments. Unfortunately, it is by now generally accepted that
it does not seem possible to get sufficient baryon asymmetry entirely within
the standard model. The requirements of a strong first order phase transition,
large CP violation, and suppression of baryon violating interactions in
the broken phase (so that produced asymmetry survives), seem only
possible if one considers extensions of the standard model. It is also known
that in the case of a second order phase transition, when the temperature 
decreases due to the expansion of the universe, the resulting baryon asymmetry
is 8 to 9 orders of magnitude below the required value \cite{scnd}. This 
happens essentially because the cooling rate of the universe, and hence the 
rate of change of the vacuum expectation value of the Higgs field
(say, in two Higgs extension of the Standard Model), is many orders of 
magnitude smaller than the electroweak scale \cite{scnd}. There have
been some proposals such as those based on topological defects 
\cite{td} (e.g., utilizing the symmetric core of the 
defect), and reheating of local regions due to evaporation of 
primordial black holes \cite{pbh1,pbh2}, where it has been argued that 
sufficient baryon asymmetry may arise even with a second order phase 
transition.

 An important aspect of phase transitions in the universe, which has
not received much attention, is the fact that the universe is necessarily
inhomogeneous. There are density fluctuations present in the universe, which 
eventually lead to the formation of structure we see today.  Density 
fluctuations, and hence temperature fluctuations, even if they are of 
small magnitude, can affect the phase transition dynamics in crucial ways. 
There has been some discussion of the effects of inhomogeneities on
the dynamics of a first order phase transition, in the context of 
quark-hadron transition in the universe \cite{impur,inhm,sheet}. For example,
Christiansen and Madsen have discussed \cite{impur} heterogeneous nucleation 
of hadronic bubbles due to presence of impurities. It is mentioned in 
ref.\cite{impur} that possible sources of such impurities could be 
primordial black holes, cosmic strings, magnetic monopoles, or relic 
fluctuations from the electroweak scale. Hadronic bubbles  are expected to 
nucleate at these impurities with enhanced rates. Recently, Ignatius and 
Schwarz have proposed \cite{inhm} that the presence of density fluctuations 
(those arising from inflation) at  quark-hadron transition  
will lead to splitting of the region in hot and cold regions 
with cold regions converting to hadronic phase first. Baryons will then 
get trapped in the (initially) hotter regions. Estimates of sizes and 
separations of such density fluctuations were made in ref.\cite{inhm} 
using COBE measurements \cite{cobe} of the temperature fluctuations
in the cosmic microwave background radiation (CMBR). In an earlier 
work \cite{sheet}, we have considered the effect of cosmic string
induced density fluctuations on quark-hadron transition and have shown that 
it can lead to formation of planar regions of baryon inhomogeneity which 
may have important effects on nucleosynthesis.

 There are many possible sources of density fluctuations in the 
early universe such as inflation and cosmic defects (strings, 
monopoles, textures, domain walls). There has been extensive 
study of density fluctuations generated by cosmic strings from 
the point of view of structure formation \cite{str1}. Though 
recent measurements of temperature anisotropies in the microwave background 
by BOOMERANG, and MAXIMA experiments \cite{expt} at angular scales of $\ell 
\simeq$ 200 disfavor models of structure formation based exclusively on 
cosmic strings \cite{str2,str3}, still, due to many uncertainties in the 
scaling models of cosmic string network evolution one can not rule them out
as candidates of sources of required density fluctuations. 
Further, even with present models, it is not ruled 
out that cosmic strings may contribute to some part in the structure 
formation in the universe. Above all, cosmic strings generically arise
in many Grand Unified Theory (GUT) models. If the GUT scale is somewhat
lower than $10^{16}$ GeV then the resulting cosmic strings will not
be relevant for structure formation (for a discussion of these issues, 
see \cite{str3}). However, they may still affect various
stages of the evolution of the universe in important ways. 

In this paper we will study the implications of density fluctuations 
produced by cosmic strings on electroweak 
baryogenesis for the case of a second 
order electroweak phase transition.  We will also briefly comment on the
first order transition case, a detailed discussion of that case is
postponed for a future work where we will discuss the effects of
inflation generated density fluctuations on electroweak baryogenesis. 

It is well known that cosmic strings can produce wake like 
over-densities  with magnitude which can be large \cite{str1}. (As we
will discuss below, this is also true for strings moving through
a relativistic ideal fluid). 
We consider the universe to be at a temperature 
$T_b$ which is below the electroweak transition temperature $T_c$, and 
consider density fluctuations via string wakes. Typical wavelengths 
associated with such wakes will be much smaller than the 
horizon (for ultrarelativistic strings). The evolution of 
such density fluctuations is simple for small amplitudes \cite{pdmn}; 
they propagate as plane waves with speed of sound $c_s = 1/\sqrt{3}$.
Even for larger amplitudes, the propagating density disturbance
generated by string wakes will typically retain short wavelength \cite{pdmn}.
At a given region, through which the fluctuation propagates, the density,
and hence the temperature, undergoes an oscillation, with the oscillation
time scale being governed by the typical wavelength of the fluctuation. As 
a propagating density disturbance will be like a pulse of short
wavelength, the temperature in a given spatial region in its path 
will typically undergo one oscillation cycle with short time period.

 Depending on the amplitude of temperature oscillation, and the background
temperature $T_b$ of the universe, it may happen that the temperature
temporarily exceeds the electroweak transition temperature $T_c$, thereby
restoring the electroweak symmetry. As the temperature oscillates back to 
values lower than $T_c$, the symmetry will be broken again spontaneously.
During this re-occurrence of the electroweak phase transition, baryon 
asymmetry will be generated. For a second order phase transition no baryons 
would have been generated when the whole universe went through the electroweak 
phase transition, due to extremely slow cooling of the universe \cite{scnd}. 
However, the cooling time scale, and hence the time scale of the phase 
transition, due to the density (temperature) fluctuation pulse will 
be governed by the wavelength of the fluctuation, and can be much shorter
than the Hubble time scale. The baryon asymmetry generated in this case,
therefore, can be enhanced by  many orders of magnitude. We will show that 
for a $10^{16}$ GeV GUT scale cosmic string, the resulting density
fluctuations can give the required value of the  baryon to entropy ratio
(in the context of a two Higgs doublet model as in \cite{scnd}, with the CP 
violation parameter of order 10) if the fluctuation propagates without
getting significantly damped. If fluctuations are dissipated
rapidly, then a volume suppression factor reduces the produced
baryon asymmetry, though it is still 3-4 orders of magnitude larger
than the conventional case where baryons are produced during the
cooling of the universe via expansion. 

 The paper is organized in the following manner. In section II  we briefly 
recall the relevant results for the electroweak baryogenesis for the
case when the electroweak phase transition is of second order. Here,
the baryon to entropy ratio is expressed in terms of ${\dot \theta}$, the 
time derivative of the (relative) Higgs phase, and consequently ${\dot T}$,
where $T$ is the temperature. Since largest baryon asymmetry will
arise from the largest value of ${\dot \theta}$, which translates to 
the requirement of largest value of ${\dot T}$,  it is clear that the 
most relevant density fluctuations  in our model are those with the 
shortest wavelength (for a given value of the amplitude of density 
fluctuation). In section III, we discuss such density fluctuations as 
expected from cosmic strings moving through a relativistic fluid. 
Section IV presents our results for the baryon to entropy ratio with the 
value of ${\dot T}$ governed by short wavelength density 
fluctuations. Section V briefly discusses the expected results when the 
transition is first order. Conclusions are presented in section VI.

\section{Baryon to Entropy Ratio with a Second Order Electroweak Transition}

 We first briefly outline the results for baryon to entropy ratio (B/s)
when the electroweak phase transition is of second order and the phase
transition proceeds by the cooling of the universe due to its expansion.
For this, we will mostly follow the discussion in ref. \cite{scnd}. As
discussed in ref. \cite{mstv}, in two-Higgs doublet extensions of the 
standard model, there are terms in the effective action for the gauge-Higgs 
sector at high temperature, which are of the form $\sim \chi F {\tilde F}$.
Here $\chi$ is some combination of the Higgs fields, $F$ is the SU(2)
field strength tensor, and ${\tilde F}$ is its dual. Integration by parts
shows (with the use of the chiral anomaly equation) that this term is 
equivalent to a chemical potential term ${\dot \chi} B$ for the baryon 
number, where $B$ is the baryon number density \cite{scnd,mstv} 
(see, also, ref. \cite{nlew}). The value of this chemical potential 
$\mu_B$ can be written as \cite{scnd,mstv},

\begin{equation}
\mu_B ~ \equiv ~ {\dot \chi} ~ \simeq ~ 7 \zeta(3) ({m_t \over \pi T})^2 
{v_2^2 \over v_1^2 + v_2^2} {\dot \theta} ,
\end{equation}

\noindent where $\zeta(3) \simeq 1.2$, and $\theta$ is the relative phase 
between the two Higgs fields. $v_1$ and $v_2$ are the magnitudes of the
vacuum expectation values of the two Higgs fields, and $m_t$ is the top 
quark mass. In the presence of
chemical potential given in Eq.(1), one can show \cite{scnd} that the 
rate at which the baryon number relaxes to equilibrium, is given by

\begin{equation}
{\dot B} = - \Gamma^\prime (B - c_n \mu_B T^2) ,
\end{equation}

\noindent where $\Gamma^\prime  = {N_F^2 \Gamma \over c_n T^3}$, and $c_n$ 
is a slowly varying parameter with value remaining close to about 0.4. 
$N_F$ is the number of quark generations, and $\Gamma$ is the sphaleron 
rate per unit volume in the broken phase \cite{sphl}. Using the 
Green's  function method, the solution to this equation is obtained as

\begin{equation}
B(t) = {\int_{t_{in}}^{t} {\Gamma^\prime(t^{\prime}) c_n \mu_B(t^{\prime})
T^2(t^\prime) [e^{-\int_{t^{\prime}}^{t} \Gamma^\prime
(t^{\prime\prime})dt^{\prime\prime}}] dt^{\prime}}} .
\end{equation}

If we make the assumption that ${\int_{t_{in}}^t} {\Gamma^\prime dt \gg 1}$,
then the upper bound of this equation is given by the maximum of the
function $ c_n \mu_B(t) T^2(t)$ which we call as $B_{0}$. 
The solution is then  simply 
 
\begin{equation}
B(t) \leq B_{0} {\int_{t_{in}}^{t} {\Gamma^\prime(t^{\prime}) 
     [e^{-\int_{t^{\prime}}^{t} {\Gamma^\prime
(t^{\prime\prime})dt^{\prime\prime}}}]dt^{\prime}}} = B_{0}
\equiv |c_n \mu_B(t) T^2(t)|_{max} . 
\end{equation}

 We will use this upper limit $B_0$ as an estimate of the value of baryon 
asymmetry created. We write $B_0$ as follows,

\begin{equation}
B_0 = |c_n {\dot \chi} T^2|_{max} = |c_n {d\chi \over dT} {\dot T}
T^2|_{max} .
\end{equation}

The value of $({d\chi \over dT})_{max}$ will crucially depend on the 
details of the phase transition dynamics. Uncertainties in these
estimates have been discussed in the literature \cite{scnd}. For simple
estimates, one can use,

\begin{equation}
T{d\chi \over dT} \simeq {d\phi \over dT}\epsilon \simeq \epsilon ,
\end{equation}

\noindent where $\epsilon$ characterizes  CP violation in the model, $\phi$
being the Higgs field. Assuming
order one CP violation, one can take $T{d\chi \over dT} \simeq 1$. It has 
been argued \cite{scnd} that depending on the details of the specific
models, this value may be larger  by about two orders of magnitude. One can
thus use $\epsilon$ as a parameter, with its value ranging from 1 to
about 200 (ref. \cite{scnd}). 
 
If the departure from equilibrium is due to the expansion of the
universe, then the Friedmann equation for the radiation era can be used 
to relate the time derivative to the temperature derivative. In the
radiation era,

\begin{equation}
H(T) = ({8 \pi^3 G g_* \over 90})^{1/2} T^2 = {1 \over 2 t} .
\end{equation}

 Here $G$ is the Newton's constant, $H^{-1}$ is the Hubble time, and we 
have used the plasma energy density $\rho = {\pi^2 g_* T^4 \over 30}$
with $g_*$ being the number of degrees of freedom. We get,

\begin{eqnarray}
\dot{T} = - H(T) T  ,\hspace{0.5cm}   H(T_{c}) = H(T) {T_{c}^2 \over T^2} .
\end{eqnarray}

Using these equations, we can write $B_0$ as,

\begin{equation}
B_0 = c_n \epsilon H(T) T^2 .
\end{equation}

The equation for the entropy density is given by  

\begin{equation}
s = {2 \pi^2 \over 45} g_{*} T_{c}^3  .
\end{equation}

 Resulting baryon to entropy ratio can now be written as,

\begin{equation}
{B \over s } \simeq {45 c_n \epsilon  \over 2\pi^2 g_*} {H(T_c) \over T_c} .
\end{equation}

For the electroweak case the value of $g_{*}$ is about 100. With $T_{c} = 
110 GeV$ and ${H \over T_c} \sim 10^{-16}$, we get,

\begin{equation}
{B \over s } \simeq 10^{-18} \epsilon .
\end{equation}

 Thus, as discussed in the literature \cite{scnd}, even with optimistic 
estimates of $\epsilon \sim 200$,  required baryon asymmetry cannot be 
generated for the case of second order transition when the transition 
proceeds by the cooling of the universe due to expansion. 

 In the next sections we will discuss the density fluctuations present
in the universe. We will see that in the presence of these fluctuations,
the time scale of phase transition can be smaller by several orders of
magnitude, such that required baryon asymmetry may be generated with
a second order transition.

\section{Density Fluctuations in the Universe}

 Results for B/s obtained in the previous section correspond to the situation
when the universe is perfectly homogeneous, and undergoes second order
phase transition everywhere as the temperature falls below $T_c$ due
to the expansion of the universe. As mentioned in the Introduction, presence
of density fluctuations in the universe is unavoidable, as then only 
structures observed today can arise. In this section we will discuss 
the nature of density fluctuations produced via cosmic strings moving 
through the cosmic fluid. For the second order transition case, 
density fluctuations produced via inflation will not be helpful in
generating a baryon asymmetry in our model, as we will explain at the
end of Sec.IV. For simplicity,
we discuss the case of straight cosmic strings which give rise to wake like
density fluctuations. It is important to note that for cosmic strings, 
another important contribution to density fluctuations comes from small 
loops. However, the exact nature of density fluctuations in this case is 
more complicated due to the fact that loops rapidly oscillate, and the 
issue of time dependence is of crucial importance for our results. In a 
more detailed investigation their contribution must be taken into account.
(Segments of very large string loops, which have sizes not much smaller 
than the horizon, will effectively behave as straight strings).

  We first discuss the evolution of such small wavelength density 
fluctuations \cite{pdmn}. It is well known that the 
density fluctuations in a relativistic ideal 
fluid which have wavelength of order, or larger, than the horizon, grow
with the scale factor. However, when wavelength of the fluctuation is
much smaller than the horizon, which is the relevant case for us, then
the density fluctuation simply propagates as a plane wave with the speed of
sound $c_s = 1/\sqrt{3}$, at least when the amplitude of density
fluctuation is small \cite{pdmn}. 
Thus, such a density fluctuation will evolve as,  

\begin{equation}
{\delta \rho \over \rho} \sim A~ e^{i({\bf k.r} - \omega t)} ,
\end{equation}

\noindent where A is the amplitude of the fluctuation, ${\bf k}$ is the 
wave vector and $\omega (= 2\pi c_s/\lambda)$ is the angular frequency of 
the fluctuation with wavelength $\lambda$. 
Our interest will be in the evolution of density
disturbances for times shorter than the Hubble time, so the effects
of universe expansion can be neglected for these fluctuations for 
an order of magnitude estimate. Also, the wake like density
fluctuation arising from cosmic string moving through a relativistic
ideal fluid (with zero chemical potential) may not lead to sustained 
oscillations. What one expects is that the resulting density
fluctuation (hence temperature fluctuation) will propagate as a 
single pulse.

  It is important to note here that the main factor responsible for
baryon production enhancement in our model will be (as we will
see below) the short time scale associated with these {\it small
scale} density fluctuations compared to the Hubble time. This 
remains true irrespective of the fact whether these fluctuations 
propagate for significant distances, or they decay rapidly. If the 
fluctuation decays without propagating significantly, still the time
scale for the temperature fluctuation will be the same as for the 
propagating case, as again it will be typically governed by the speed 
of sound $c_s$ and the length scale of the fluctuation. As we 
will see below, for the first case, the propagating disturbances
can lead to required baryon production as they sweep the entire
horizon volume. If fluctuations decay rapidly, then there will be
a volume suppression factor which will lead to much smaller
baryon production, though it will still be several orders of 
magnitude larger than the conventional case when baryons are produced
only due to the overall cooling of the universe. 

  Once the nature of density fluctuations is known (depending on the source 
of density fluctuations), we can use Eq.(13) to determine the time variation 
of temperature resulting from such propagating density fluctuations. An
important assumption we make in this regard is that temperature at any
time and at any point of space, is determined by the local density $\rho$,
even in the presence of short wavelength density fluctuations. 
This assumption of local thermal equilibrium is justified here because
the wavelengths we will be considering will be about $10^{-6}$ cm,
with the relevant time scale of density variation being of order $10^{-16}$
sec. This time scale, though much shorter than the Hubble time, is still 
many orders of magnitude larger than the time scale of relevant interactions, 
which keep the particles in equilibrium. 
 
 With the use of local thermal equilibrium assumption, the density $\rho$ is 
related to the temperature $T$ as, $\rho = {\pi^2 g_* \over 30} T^4$, where
$g_* \simeq 100$ is the number of degrees of freedom at the electroweak
scale.  From this we can relate the temperature fluctuation to the
density fluctuation,

\begin{equation}
{\delta\rho \over \rho} \equiv {{\rho - \rho_{b}} \over \rho_{b} } =  
{{T^4 - {T_{b}}^4} \over {T_{b}}^4 } ,
\end{equation}

\noindent where $T_{b}$ is the background temperature of the plasma. Then 
using Eq.(13) we obtain the time - temperature relation for propagating 
density fluctuations.

\begin{equation}
{{T^4 - {T_{b}}^4} \over {T_{b}}^4 } = A~ e^{i({\bf k.r} - \omega t)} .
\end{equation}

 Given this, we take the time dependence of the temperature at a given 
point in space to be given by,

\begin{equation}
T(t) = T_{b} [ 1 + A~ \sin{\omega t}]^{1/4} .
\end{equation} 

 Thus, at $t = 0$, $T = T_{b}$, the background temperature.

  As discussed above, although we are using the above
expression for $T(t)$ with a periodic time dependence, the propagating
density fluctuation actually constitutes only a single pulse. Thus the
above expression for $T(t)$ is to be used only for a single cycle.
Of course that is enough for our purpose as the baryon asymmetry is to 
be generated during the cooling part of the cycle, irrespective of
the number of cycles.

 Propagation of density fluctuation given in Eq.(13), and resulting
oscillation of temperature (Eq.(16)) is valid for the case when
the amplitude $A$ is small, so the disturbance propagates as acoustic
wave. For our case, density fluctuations produced in string wakes can have
large amplitudes, with $A \sim 1$. Evolution of such density fluctuations 
is not as simple as given above. In fact, for a general fluctuation (i.e.
with large amplitude), the propagating
disturbance can lead to transfer of energy to shorter wavelengths,
leading to steepening of the disturbance \cite{pdmn}. In our case
the initial profile of the density fluctuation will be given by the
wake like shock which forms behind a cosmic string moving with 
supersonic velocity. Also, the baryon asymmetry produced in our model will
be larger for shorter wavelengths due to shorter time scale of
temperature variation. Thus, a propagating disturbance even with
large amplitude will lead to baryon production at least as large
as given by the initial wavelength of the disturbance. This is 
assuming that the disturbance does not decay away very rapidly, e.g.
by subsequent shock developments etc. Though, as we have discussed
above, even if fluctuation decays rapidly, the typical time scale of 
temperature evolution will still be given by the wavelength of the 
fluctuations. With this in mind, we will use the temperature
variation given by Eq.(16) (again, for one cycle only as appropriate
for a pulse of over-density) even for the case when the amplitude $A$ of 
density fluctuation is not too small, as in the case of a relativistic
cosmic string wake.

\subsection{Density Fluctuations arising due to Straight Cosmic Strings}

 We now give a brief review of the structure of density 
fluctuations produced by a cosmic string moving through a relativistic fluid.
The space-time around a straight cosmic string (along the z axis) is given 
by the following metric \cite{mtrc},

\begin{equation}
ds^2 = dt^2 - dz^2 - dr^2 -  (1 - 4G\mu)^2 r^2 d\psi^2 ,
\end{equation}

\noindent where $\mu$ is the string tension. This metric describes a conical 
space, with a deficit angle of $8\pi G\mu $.  This metric can be put in the 
form of the Minkowski metric by defining angle $\psi^\prime = (1-4G\mu) 
\psi$. However, now $\psi^\prime$ varies between 0 and $(1-4G\mu)2\pi$, that 
is, a wedge of opening angle $8\pi G\mu$ is removed from the Minkowski 
space, with the two boundaries of the wedge being identified. It is well 
known that in this space-time, two geodesics going along the opposite sides 
of the string, bend towards each other \cite{gdsk}. This results in binary 
images of distant objects, and can lead to planar density fluctuations. These
wakes arise as the string moves through the background medium, giving
rise to velocity impulse for the particles in the direction of the 
surface swept by the moving string. For collisionless cold dark matter
particles the resulting velocity impulse is \cite{str1,wake}, $v_{impls} 
\simeq 4\pi G\mu v_{st} \gamma_{st}$ (where $v_{st}$ is the transverse 
velocity of the string).  This leads to a wedge like region of overdensity, 
with the wedge angle being of order of the deficit angle, i.e. $8\pi G\mu$
($\sim 10^{-5}$ for 10$^{16}$ GUT strings). The density fluctuation in the 
wake is of order one. Subsequent growth of this wake by gravitational 
instability in the matter dominated era has been analyzed in great detail in 
literature \cite{wake}. 

The structure of this wake is easy to see for collisionless particles 
(whether non-relativistic, or relativistic). Each particle trajectory 
passing by the string bends by an angle of order $4\pi G\mu$ towards the 
string. In the string rest frame, take the string to be at the origin, 
aligned along the z axis, such that the 
particles are moving along the $-x$ axis. Then it is easy to see that 
particles coming from positive $x$ axis in the upper/lower half plane will  
all be above/below the line making an angle $\mp 4\pi G\mu$ from the 
negative $x$ axis. This implies that the particles will overlap in the 
wedge of angle $8\pi G\mu$ behind the string leading to a wake
with density twice of the background density. One thus 
expects a wake with half angle $\theta_w$ and an overdensity
$\delta \rho/\rho$ where \cite{str1,wake},

\begin{equation}
\theta_w \sim 4\pi G\mu, \qquad {\delta \rho \over \rho} \sim 1 .
\end{equation}

 However, the case of relevance for us is cosmic strings moving through
a relativistic plasma of elementary particles at temperatures of order 
100 GeV.  At that stage, it is not proper to take the matter as consisting
of collisionless particles. A suitable description of matter at that
stage is in terms of a relativistic fluid which we will take to be
an ideal fluid consisting of elementary particles.
Generation of density fluctuations due to a cosmic string moving 
through a relativistic fluid has been analyzed in the literature
\cite{shk1,shk2,shk3}. The study in ref.\cite{shk1} focused on the
properties of shock formed due to supersonic motion of the string 
through the fluid. In the weak shock approximation, one finds a 
wake of overdensity behind the string. In this treatment one can
not get very strong shocks with large overdensities. In refs.
\cite{shk2,shk3}, a general relativistic treatment of the shock
was given which is also applicable for ultra-relativistic string
velocities. The treatment in ref.\cite{shk3} is more complete
in the sense that the equations of motion of a relativistic fluid 
are solved in the string space-time (Eq.(17)), and both subsonic 
and supersonic flows are analyzed. One finds that for  
supersonic flow, a shock develops behind the string, just as 
in the study of ref. \cite{shk1,shk2}. In the treatment of 
ref.\cite{shk3} one recovers the usual wake structure of overdensity 
(with the wake angle being of order $G\mu$) as the string approaches 
ultra-relativistic velocities. Also the overdensity becomes of order 
one in this regime.

 We mention here that it is not expected that the 
string will move with ultra-relativistic velocities in the early universe. 
Various simulations have shown \cite{vstr} that rms velocity of string 
segments is about 0.6 for which the shock will be weak. For this case,
the opening angle of the wedge $\theta_w$ will also be large. This
will then imply a large wavelength for the density fluctuation, and
consecutively, small value of ${\dot T}$ and small baryon to entropy ratio.
It is still possible that various segments of string may move
with ultra-relativistic velocities. (Note that the friction dominated
regime for GUT strings ends long before these temperatures are reached
\cite{friction}.) We will take this to be the case, and will consider
the situation when string produces strong shock, with a small wedge angle.

 For ultra-relativistic case, we  use expressions from ref.\cite{shk3}. 
Resulting density fluctuation in the wake of the moving string is
expressed in terms of fluid and sound four velocities,

\begin{equation}
{\delta\rho \over \rho} \simeq {16\pi G\mu u_f^2(1+u_s^2) \over 
3u_s\sqrt{u_f^2 - u_s^2}}, \qquad sin\theta_w \simeq {u_s \over u_f} ,
\end{equation}

\noindent where $u_f = v_f/\sqrt{1-v_f^2}$ and $u_s = c_s/\sqrt{1-c_s^2}$. 
In this case, when string velocity $v_f$ is ultra-relativistic, then
one can get strong overdensities (of order 1) and the angle of the
wake approaches the deficit angle $\simeq 8\pi G\mu$. This is the structure 
of wake which is also expected for a collision-less matter as discussed 
above. We will assume that this is the case, and hence, take the properties 
of the shock as given by Eq.(18). The density fluctuations resulting from 
(straight) cosmic strings, as characterized by Eq.(18), have the amplitude 
$A$ of order 1, and average wavelength $\lambda$ of order (1/2)($8 
\pi G \mu d_H$) where $d_H$ is the horizon size. For a GUT scale cosmic 
string with $8\pi G \mu \sim 10^{-5}$, $\lambda$ is of order $10^{-6}$ 
cm. Note that this value of $\lambda$ is much larger than the
diffusion length scales $l_{diff}$ of various particles at these 
temperatures which is of order 10$^{-14}$ cm \cite{neutr}. Hence 
these fluctuations will not be dissipated by diffusion processes.

\section{Baryon to Entropy ratio due to Density Fluctuations}

 In the previous section, we have determined the time variation
of the temperature resulting from density fluctuations. From 
Eqs.(5),(6), and (16), we can write,

\begin{equation}
B_0 = (c_n \epsilon T {\dot T})_{max} = {c_n  \epsilon A \omega
T_b^4 \over  4 T^2} .
\end{equation}

 This estimate of maximum baryon asymmetry has been obtained with the
optimistic assumption that sphaleron processes turn off when the source 
of baryon number has optimum value \cite{scnd}. As the sphaleron rate
changes by many orders of magnitude when $T$ decreases below $T_c$, relevant 
value of $T$ in the above equation will be close to $T_c$. For rough
estimates, we will take $T \sim T_b$ (note that $T_b$ is of same order
as $T_c$ in our model as we want density fluctuations to
be able to restore the electroweak symmetry locally). With this, Eq.(20)
is obtained with maximum value of $cos\omega t$ = 1.
The resulting baryon to entropy ratio can now be  expressed 
in terms of the amplitude $A$, and the wavelength $\lambda$ of the 
density fluctuation as,

\begin{equation}
{B \over s} ~ = ~ {45 c_n c_s \over \pi g_*} {A \over 4 \lambda T_b} 
\epsilon ~ \simeq ~ {0.01 \epsilon A \over \lambda T_b} . 
\end{equation}

Here we have used $g_* = 100$. 

For density fluctuations generated due to cosmic string wakes, the
amplitude $A \sim 1$, and  the  shortest wavelength is given by the 
average width of the generated wake $ \sim {1 \over 2} 8 \pi G \mu d_H $ 
(since the wake has the shape of a wedge) where $d_{H}$ is the horizon 
size at that time. Since $ d_{H}$ is of the order of 0.1 cm at the time 
of the electroweak scale, and $ 8\pi G \mu \simeq 10^{-5} $ for  
a GUT string, we get $\lambda_{csmc} \simeq 10^{-6} cm $. The resulting value 
of baryon to entropy ratio from Eq.(21) is $B/s \simeq 10^{-11} \epsilon$. 
We see that for the case of cosmic string, with $\epsilon$ of order 10,
one is able to get the required baryon asymmetry. (We mention here that with 
the amplitude of density fluctuations of order one, $T_b$ can be much 
lower than $T_c$ in the cosmic string case. Thus, there may be possibility 
of having a weaker bound on the Higgs mass which is needed to avoid sphaleron 
wash out of the created baryon asymmetry, see ref.\cite{pbh2}.)

 It is important to realize that, if fluctuations propagate without
getting significantly damped, then
essentially all regions will participate in 
this generation of baryon asymmetry, so there is no volume suppression
factor here. For example, the structure of density fluctuations here
is in form of planar sheets (wakes). These planar
density fluctuations will propagate in the direction normal to the plane, and 
sweep the entire horizon volume. At any given point within the horizon volume,
the temperature will undergo an oscillation as given by Eq.(16). There may 
be some horizon volumes which will not have any strings present to generate  
significant density fluctuations at the relevant stage. However, the
probability of occurrence of such regions will be small, and hence the order 
of magnitude estimate of the baryon to entropy ratio as given above will
not be affected. There will be some change in the wavelengths 
of these fluctuations during the time they propagate across 
the entire horizon, but again, one will expect it not to affect 
the final numbers within an order of magnitude. Some level of 
dissipation of density fluctuations, while they 
propagate, may be compensated by the fact that there will be
multiple sources of density fluctuations in a given horizon volume. For
example, string simulations have shown \cite{stntwrk} that the number of 
long strings in a given horizon volume is expected to be about 15. Each
of these strings will lead to wakes of density fluctuations, which will
sweep the horizon volume. This may also compensate for the effects of
shorter wakes of a given strings (possibly getting truncated behind
the string due to other strings/fluctuations). It is also possible
that the baryon asymmetry generated due to one
string may first get wiped out when temperature there rises above $T_c$
due to another string wake. In that region, the surviving baryon asymmetry
will correspond to the density fluctuations which are last to pass by.
Eventually, the temperature in that region will be sufficiently low so
that any subsequent cosmic string wakes are not able to re-heat the region
enough to destroy the already created baryon asymmetry. This will be the stage
when baryon asymmetry will get frozen. 

 From above discussion we see that the most important factor responsible for
the enhancement of baryon to entropy ratio in our model is the short
time scale of these density fluctuations. In addition, propagation of
these over-density pulses leads to baryon production in almost
entire region (say in a given horizon volume). That is, there is
no volume suppression factor here. We now discuss the possibility
if these fluctuations do not propagate for large distances, and decay
very rapidly. In such a situation, to make a conservative estimate, we 
can take the over-dense wake of each string as a fixed region in space 
where the overdensity will dissipate typically with speed of sound $c_s$.
This means that the time scale of temperature variation in this wake
will again be the same as discussed above. From this we conclude that
the resulting baryon to entropy ratio in this case will be suppressed
by a volume factor $f_s$ compared to the value given in Eq.(21).
$f_s$ will be given by the ratio of the total volume inside the wakes
of all strings in a given horizon volume to the horizon volume.
Taking account of about 10 long strings in a horizon volume, with
typical wake thickness of order 10$^{-5} d_H$, we get,

\begin{equation}
f_s = {10 \times 10^{-5} d_H \times d_H^2 \over d_H^3} ~ \simeq 10^{-4} 
\end{equation}

 With various factors taken as earlier, we get net $B/s \simeq 10^{-15}
\epsilon$ for this case. With optimistic value of $\epsilon \sim 100$,
resulting value of $B/s$ is $10^{-13}$, about three orders of magnitude
smaller than the required one. There may be a possibility of further
enhancement in this value by changing various parameters of cosmic
string network e.g. string scale and string velocity (which determine the
parameters of string wakes), as well as by properly taking into 
account the effect of density fluctuations produced by all
strings at that stage, i.e. including contributions of all string
loops as well, integrated over a suitable duration of time. The main
point we would like to make is that even in this case where density
fluctuations do not propagate, our model leads
to the enhancement of baryon to entropy ratio by at least 3 - 4 orders
of magnitude compared to the case when baryogenesis happens only due
to the overall cooling of the universe.
 
 An important assumption in deriving equations such as Eq.(5),(6) (using 
which Eq.(20) has been derived) is that 
at any stage during the variation of the temperature, the value of the Higgs
field is essentially given by the vacuum expectation value ($vev$) at that
temperature. This assumption requires that the time scale of
temperature variation should be much larger than the typical time scale
of the Higgs field evolution. More importantly, as discussed above,
as the temperature rises above $T_c$ (Eq.(16)), the Higgs field should
settle down to value zero so that the electroweak symmetry is restored. 
This is very important as during the rise of temperature,
one will get production of antibaryons via above equations (as sign
of ${\dot \chi}$ will be reversed during this stage). However, if the
electroweak symmetry is restored after that, then all these net antibaryon 
asymmetry will be erased due to unsuppressed sphaleron processes in the 
symmetric phase.  Thus, only the net baryon asymmetry will remain which 
will be generated during the stage when the temperature is decreasing.

 For checking this requirement on the time scale of temperature variation, 
we take a simple effective potential to model the second order phase 
transition (as in ref.\cite{brnd}),

\begin{equation}
V(\phi, T) = D (T^2 - T_{c}^2) \phi^2 + {\lambda \over 4} \phi^4 .
\end{equation}

 We take the values of the parameters as suitable for the Standard
model\cite{brnd}, $T_c = 110$ GeV, $D \simeq 0.18$ and $\lambda \simeq 0.1$. 
The evolution equation of $\phi$ (neglecting spatial dependence) is,

 \begin{equation}
\ddot\phi + {3 \over 2}{\dot\phi \over t} + V^{\prime}(\phi,T(t)) = 0 ,
\end{equation}

\noindent where the time dependence of $T$ in $V$ is given by Eq.(16).
Using this equation, we have checked the evolution of $\phi$ for one
oscillation period, starting from the initial time $t = 0$ in Eq.(16), 
which corresponds to $T = T_b$ (the background temperature)
initially. Top figure in Fig.1 shows the time variation of the 
temperature resulting from cosmic string wakes. 
Here we have taken $T_b = 100$ GeV, $A$ = 0.8 in Eq.(16), and 
$\omega$ corresponds to $\lambda = 4 \pi G 
\mu d_H \sim 10^{-6}$ cm. The values of $T_b$ and $A$ are taken
as sample values such that  the maximum value of $T$ (Eq.(16)) reaches 
slightly above $T_c = 110 GeV$. Bottom figure in Fig.1 shows plot of the 
evolution of the Higgs field $\phi$ (solid line), and the $vev$ of $\phi$ 
(shown by solid dots at few values of $t$). Note that the plot of 
${\Delta T \over T}$ deviates
from purely sinusoidal behavior due to large value of $A$.

\begin{figure}[h]
\begin{center}
\leavevmode
\epsfysize=12truecm \vbox{\epsfbox{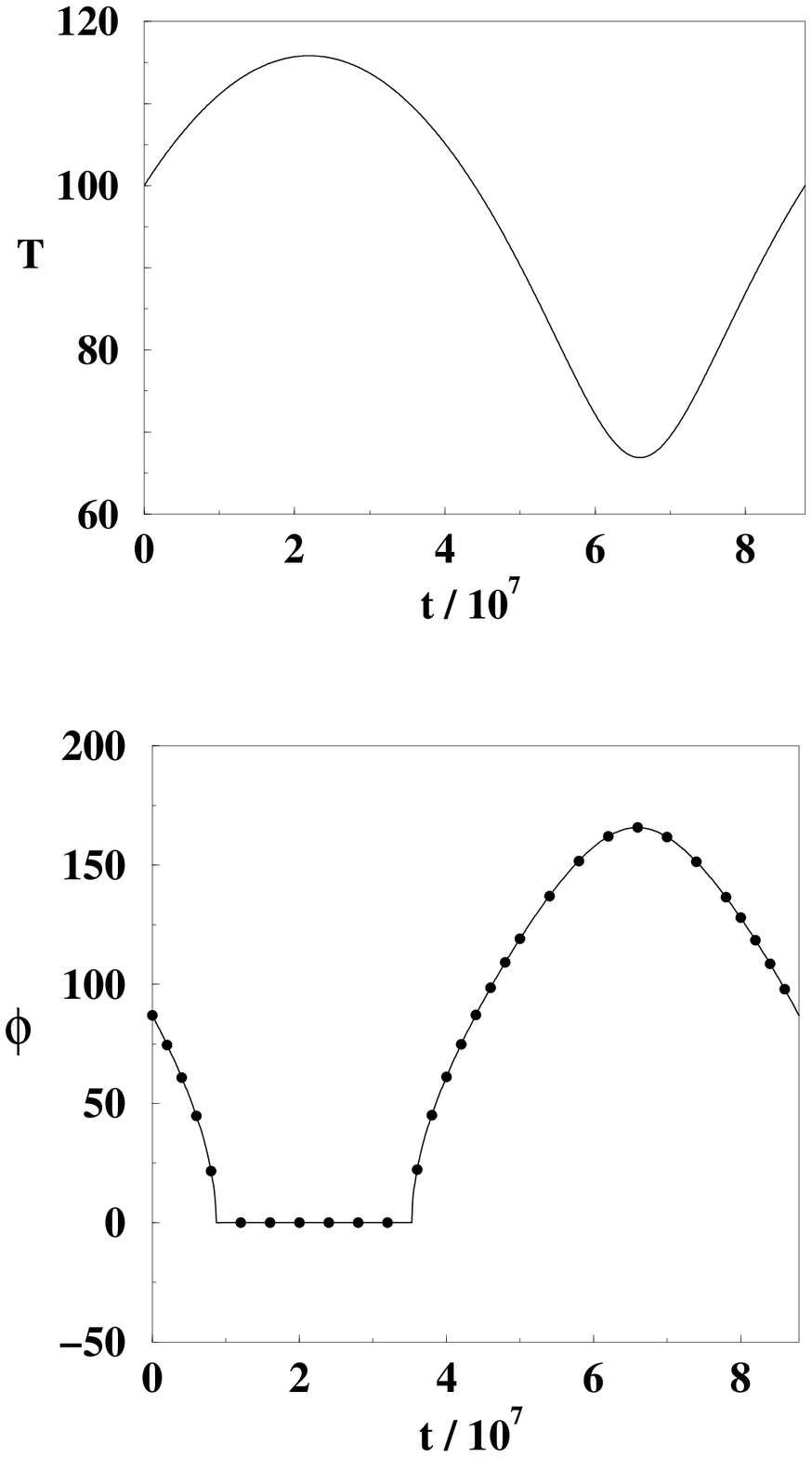}}
\end{center}
\caption{}{Top figure shows time variation of the temperature $T$ (in GeV) 
for one oscillation period, resulting from density fluctuations with 
wavelength $\lambda \sim 4\pi G \mu d_H$. Time $t$ is in $GeV^{-1}$.
Bottom figure shows the evolution of the Higgs field $\phi$ (solid line), 
and its $vev$ (dots), both in GeV.}
\label{Fig.1}
\end{figure}

 From Fig.(1) we see that $\phi$ traces the evolution of the $vev$ 
faithfully. For time duration when $T > T_c$, $\phi$ settles down 
to the value zero, thereby restoring the electroweak symmetry. 
This restoration of symmetry happens for the values of $t$ 
between $0.9 \times 10^7$ to $3.5 \times 10^7$ GeV$^{-1}$. 
As we mentioned 
above, this temporary symmetry restoration is necessary to wipe out 
the antibaryons which would have been generated during the rising phase of 
the temperature. It is also important to note from these figures that 
the time duration for which the symmetric phase lasts (i.e. when $T > T_c$)
is of the order of $10^7 GeV^{-1}$, 
which is much larger than the typical time scale of a sphaleron 
transition in the symmetric phase\cite{sphl2} which is roughly 
of order $\sim (\alpha_w^2 T)^{-1} 
\simeq  10 GeV^{-1}$. (Spatial extent of regions with density 
fluctuation is also many orders of magnitude larger than the sphaleron 
size $\sim (\alpha_w T)^{-1}$.) Thus, equilibrium calculations
used in this paper make sense, sphaleron transitions 
during the time when $T > T_c$, should be able to erase any antibaryon 
asymmetry generated earlier during the rise of temperature.

 We now comment on the density fluctuations generated via inflation. Generic
prediction of most inflationary models \cite{infl} is that the resulting 
density fluctuations are scale invariant, with a magnitude of about $10^{-4}$
(when the fluctuation re-enters the horizon), as constrained by COBE
observations of temperature fluctuations ${\delta T \over T} \sim 10^{-5}$.
In the context of our model, inflationary density fluctuations may
contribute to baryon production only if such small temperature 
increase can lead to significant change in the baryon violation rates
via sphaleron transitions. This requires that the background
temperature of the universe $T_b$ be very close to the electroweak 
transition temperature $T_c$. However, in such a situations the
produced baryon asymmetry will be erased as the sphaleron transitions
at $T = T_b$ will not be much suppressed. Basically, for the scales
involved, a fractional temperature difference of order $10^{-5}$ is
unlikely to lead to large change in sphaleron rates. However, the
situation is entirely different or the case of a first order
transition, where even very small temperature differences can lead to
significant changes in the bubble nucleation probability. We briefly
discuss it in the next section.   

\section{First Order Transition Case}

 Our conclusions about the second order transition should be
applicable in the case of a very weak first order transition (or
a cross-over). For example, after a rapid production of bubbles of
the broken phase, the bubbles will collide and the transition will be
completed. Presence of propagating density fluctuations will again reheat 
these regions temporarily, restoring the electroweak symmetry. Subsequent
cooling will then produce a net baryon asymmetry as described above,
with the rate of change of temperature governed by the oscillation
frequency of the density fluctuation.

 If the first order transition is not very weak, then the effect of density
fluctuations should manifest in entirely different manner. Now, the presence
of density (and hence) temperature inhomogeneities will split the region
into cold regions and hot regions (at a given time). Geometry of such
regions will depend on the source of density fluctuations. For example,
for inflationary density fluctuations, such regions can be taken as localized
overdense, or underdense regions, as discussed in ref. \cite{inhm}. 
Phase transition (say, via bubble
nucleation) will first happen in the colder regions,
forming large interfaces which then extend into the hotter
regions. Thus, even for the case of inflationary density fluctuations,
the process of electroweak baryogenesis may get affected significantly.
For example, it is possible that even for the case of a weak first
order transition, different temperature zones may lead to propagating
interfaces emanating from the colder regions and extending into
the hotter regions. This in effect may lead to a picture similar to
that of a strong first order transition which is one of the
requirements for a successful baryogenesis model.

Also, the propagation of the interface will reheat the surrounding region.
Due to temperature variations, density of baryons produced in this case may 
then vary from cold  regions to the hot regions. Similarly, for the case of 
cosmic strings, wedge shaped regions will form where the temperature
will be higher than the surrounding (as discussed in ref.\cite{sheet}). 
Bubble nucleation and coalescence may happen in the colder regions first, 
with the interface then extending into the planar hotter regions. This may 
lead to variations in the baryon density of planar nature. It will be
very interesting to find whether any such baryon inhomogeneities can 
survive until nucleosynthesis to affect elemental abundances.   
 
\section{Conclusion}

 We have studied the implications of small wavelength density fluctuations 
produced by cosmic strings on electroweak 
baryogenesis. We have shown that due to rapid changes in the temperature
arising from short wavelength density fluctuations, electroweak
symmetry may get restored locally, depending on the amplitude of 
temperature oscillation and the background temperature $T_b$ of the 
universe. As the temperature oscillates back to a value below the 
transition temperature, electroweak symmetry will be broken again
spontaneously. During this re-occurrence of the electroweak phase 
transition, baryon asymmetry will be generated. The cooling time scale, 
and hence the time scale of the phase 
transition, due to such density (temperature) fluctuation waves will 
be governed by the wavelength of the fluctuation, and can be much shorter
than the Hubble time scale. The baryon asymmetry generated in this case
is therefore enhanced by  many orders of magnitude. If these short
wavelength density fluctuations propagate for significant distances
without decaying, then baryogenesis happens in every region of space,
as different regions are swept by the propagating over-density (hence
temperature) pulse. We have shown that 
for a $10^{16}$ GeV GUT scale cosmic string, the resulting density
fluctuations can give the required value of the  baryon to entropy ratio
(in the context of a two Higgs doublet model as in \cite{scnd}, with the CP 
violation parameter of about 10). On the other hand, if density
fluctuations decay away rapidly, then there is a volume suppression 
factor, and the resulting baryon asymmetry is smaller by a factor
of order 10$^{-4}$ (though it is still larger by at least 3 - 4 orders
of magnitude compared to the conventional case where baryons are
produced when the universe cools via expansion).
  
 Many aspects of our model need to be worked out in more detail. Exact
nature of propagation and decay of these density perturbations needs
to be worked out. We have
used order of magnitude estimates for the chemical potential for baryons
(Eq.(6)), more careful estimates need to be worked out. Further, the 
baryon asymmetry should be
calculated by taking account of several propagating density fluctuation
waves. Net asymmetry should be calculated by detailed accounting for the
generation of antibaryon asymmetry during the period when $T$ rises, its
diffusion outside the region of density fluctuation, and the decay of 
this antibaryon asymmetry when $T \ge T_c$ so that the 
sphaleron transitions are unsuppressed. Finally, the first order 
transition case may show many interesting possibilities, especially
for the case of inflationary density fluctuations. 

Similar calculations, as reported in this paper, can be
done for other sources of density fluctuations, such as those generated
by textures etc. We expect similar results for the baryon asymmetry in
those cases. For example, for the case of textures, collapsing textures 
(at any stage) will give rise to small scale density fluctuations, 
which will then propagate as density waves. The amplitude of
fluctuations  may be high for the case of textures also since the relevant 
fluctuations will be generated during the last stages of texture
collapse, though the volume factor will lead to an overall suppression
of baryon production. 
 
\vskip .2in
\centerline {\bf ACKNOWLEDGEMENTS}
\vskip .1in

  We are very thankful to Sanatan Digal, Sayan Kar, Rajarshi Ray,
and Supratim Sengupta for useful comments. We especially thank
Raghavan Rangarajan for detailed comments and many useful suggestions.



\begin{thebibliography}{99}

\bibitem{ewb}  For a review see, A.G. Cohen, D.B. Kaplan
and A.E. Nelson, Annu. Rev. Nucl. Part. Sci. {\bf 43} (1993) 27.
For more recent developments see M. Trodden, Rev. Mod. Phys. {\bf 71}
(1999) 1463.

\bibitem{scnd} M. Joyce and T. Prokopec, Phys. Rev. {\bf D57}, 6022 (1998). 

\bibitem{td} R. H. Brandenberger, A.C. Davis, M. Hindmarsh, 
Phys. Lett. {\bf B263}, 239 (1991); R. Brandenberger, A.C. Davis,
Mark Trodden, Phys. Lett. {\bf B335}, 123 (1994); 
R. Brandenberger, A.C. Davis, T. Prokopec, and M. Trodden, 
Phys. Rev. {\bf D53}, 4257 (1996); T. Prokopec, R.  Brandenberger, 
A.C. Davis, and M. Trodden, Phys. Lett. {\bf B384}, 175 (1996);
R. Brandenberger and A. Riotto, Phys. Lett. {\bf B445}, 323 (1999).

\bibitem{pbh1} Y. Nagatani, Phys. Rev. D {\bf59}, 041301 (1999), see also,
hep-ph/0104160.

\bibitem{pbh2} R. Rangarajan, S. Sengupta, and A.M. Srivastava,
hep-ph/9911488, Astropart. Phys. {\bf 17}, 167 (2002).

\bibitem{impur} M.B. Christiansen and J. Madsen, Phys. Rev.
{\bf D53}, 5446 (1996).
 
\bibitem{inhm} J. Ignatius and D.J. Schwarz, Phys. Rev. Lett.
{\bf 86}, 2216 (2001). 

\bibitem{sheet} B. Layek, S. Sanyal and A.M. Srivastava, 
Phys. Rev. {\bf D63}, 083512 (2001).

\bibitem{cobe} C.L. Bennett et al., Astrophys. J. {\bf 464}, L1 (1996).

\bibitem{str1} A.Vilenkin and E.P.S.Shellard, ``Cosmic 
Strings and Other Topological Defects", (Cambridge University Press, 
Cambridge, 1994). L. Perivolaropoulos, astro-ph/9410097.

\bibitem{expt} A. Lange et al., Phys. Rev. {\bf D63}, 042001 (2001); 
A. Balbi et al., Astrophys. J. {\bf 545}, L1 (2000). 

\bibitem{str2} R. Durrer, astro-ph/0003363; C.R. Contaldi, 
astro-ph/0005115; L. Pogosian, astro-ph/0009307.

\bibitem{str3} A. Albrecht, astro-ph/0009129.

\bibitem{pdmn} P.J.E. Peebles, {\it The large-scale structure of the 
universe}, (Princeton University Press, Princeton, NJ, 1980);
T. Padmanabhan, {\it Structure formation in the universe},
(Cambridge University Press, Cambridge, 1993);
L. Landau and E. Lifshitz, {\it Fluid Mechanics}
(Pergamon Press Ltd., London, 1959).

\bibitem{mstv} L. McLerran, M. Shaposhnikov, N. Turok, and
M. Voloshin, Phys. Lett. {\bf B256}, 451 (1991).

\bibitem{nlew} M.Joyce, T. Prokopec, and N. Turok, Phys.Rev.{\bf D53}, 
2930 (1996); {\it ibid},  {\bf D53}, 2958 (1996); J.M. Cline,
K. Kainulainen, and A.P. Vischer, Phys. Rev. {\bf D54}, 2451 (1996);
N. Turok and J. Zadrozny, Nucl. Phys. {\bf B358}, 471 (1991).

\bibitem{sphl} P. Arnold and L. McLerran, Phys. Rev. {\bf D36}, 581 (1987);
{\it ibid}, {\bf D37}, 1020 (1988).  

\bibitem{mtrc} J.R. Gott III, Astrophys. J. {\bf 288}, 422 (1985);
W.A. Hiscock, Phys. Rev. {\bf D31}, 3288 (1985).

\bibitem{gdsk} D.V. Gal'tsov and E. Masar, Class. Quant. Grav.{\bf 6},
1313, (1989) 

\bibitem{wake} A. Sornborger, Phys. Rev. {\bf D56}, 6139 (1997). 

\bibitem{shk1} A. Stebbins, S. Veeraraghavan, R. Brandenberger,
J. Silk, and N. Turok, Astrophys. J. {\bf 322}, 1 (1987).

\bibitem{shk2} N. Deruelle and B. Linet, Class. Quant. Grav.
{\bf 5}, 55 (1988).

\bibitem{shk3} W.A. Hiscock and J.B. Lail, Phys. Rev. {\bf D37},
869 (1988).

\bibitem{vstr} D.P. Bennett and F.R. Bouchet, Phys. Rev. {\bf D41},
2408 (1990). 

\bibitem{friction} C.J.A.P. Martins and E.P.S. Shellard, 
Phys. Rev. {\bf D53}, R575 (1996).

\bibitem{neutr} K. Jedamzik and G.M. Fuller, Astrophys. J. {\bf 423},
33 (1994). 

\bibitem{stntwrk} B. Allen and E.P.S. Shellard, Phys. Rev. Lett.
{\bf 64}, 119 (1990). 

\bibitem{brnd} T. Prokopec, R. H. Brandenberger, and A. Davis, 
hep-ph/9601327.

\bibitem{sphl2} G.D. Moore, Phys. Rev. {\bf D62}, 085011 (2000);
P. Arnold, D. Son, and L.G. Yaffe, Phys. Rev. {\bf D55}, 6264 (1997).

\bibitem{infl} For reviews, see, A. Albrecht, astro-ph/0007247;
K.A. Olive, Phys. Rept. {\bf 190}, 307 (1990).

\end{thebibliography}
\end{document}